\documentclass[amsmath,amssymb,twocolumn]{revtex4}
\usepackage{graphicx}% Include figure files

\let\^=\hat

\begin{document}
\title{Elimination of transverse instability in stripe solitons
by one-dimensional lattices}
\author{Jianke Yang$^{1,*}$, Daniel Gallardo$^{2}$, Alexandra Miller$^{2}$, and Zhigang Chen$^{2}$}
\address{$^{1}$ Department of Mathematics and Statistics, University of
Vermont, Burlington, VT 05401, USA \\
$^{2}$ Department of Physics and Astronomy, San Francisco State
University, San Francisco, CA 94132, USA \\
$^*$Corresponding author: jyang@math.uvm.edu
}%

\begin{abstract}
We demonstrate theoretically and experimentally that the transverse
instability of coherent soliton stripes can be greatly suppressed or
totally eliminated when the soliton stripes propagate in a
one-dimensional photonic lattice under self-defocusing nonlinearity.
\end{abstract}

\maketitle

\noindent It is well known that in homogeneous nonlinear optical
media, a bright soliton stripe, uniform along the transverse stripe
(say $y$) direction but localized along the orthogonal transverse
(say $x$) direction, is unstable upon propagation along the
longitudinal $z$ direction when transverse perturbations are present
\cite{Zak_Rub,TI_saturable,Saffman,Chen_Manakov,SHG_TI_neck,SHG_TI_snake,Kiv_Peli,snake,Yang_SIAM}.
When a one-dimensional (1D) optical lattice is introduced along the
$x$ or $y$ direction, the soliton stripe is still transversely
unstable under self-focusing nonlinearity
\cite{Aceves_semi_TI,1Dlattice_TI}. To suppress this transverse
instability, some ideas have been proposed. For instance, this
instability can be completely eliminated if the soliton stripe is
made sufficiently incoherent along the transverse direction
\cite{Segev}. This instability can also be significantly reduced by
nonlinearity saturation or incoherent mode coupling
\cite{Segev_Kivshar,Mull_Yang_TI}. The introduction of 2D square
lattices can also suppress transverse instability, but the
transversely stable structures in such media are soliton trains
(comprising an infinite array of intensity peaks) rather than
soliton stripes \cite{2Dtrain,2Dtrain_stable}.

In this paper, we demonstrate, both theoretically and
experimentally, that transverse instability of coherent soliton
stripes is greatly suppressed or totally eliminated when the soliton
stripes propagate in a 1D lattice under self-defocusing
nonlinearity.

\begin{figure}[h]
\includegraphics[width=0.48\textwidth]{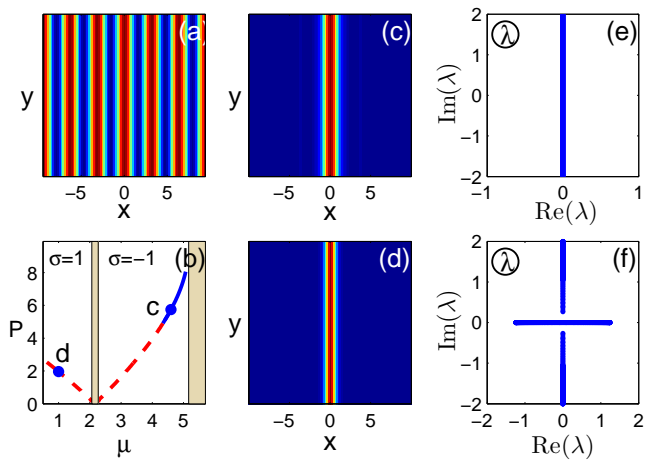}

\vspace{-0.2cm} \caption{(a) 1D lattice; (b) power curves of stripe
solitons (dashed red indicates instability, solid blue indicates
stability, and shaded regions are Bloch bands); (c) intensity
profile of a stripe soliton in the first gap (at $\mu=4.5$) under
defocusing nonlinearity; (d) intensity profile of a stripe soliton
in the semi-infinite gap (at $\mu=1$) under focusing nonlinearity;
(e,f) stability spectra of stripe solitons in (c,d) respectively. }
\label{f:fig1}
\end{figure}

Our theoretical model is the 2D NLS equation with a 1D lattice,
\begin{equation}  \label{e:model}
iU_z+U_{xx}+U_{yy}+n(x) U+\sigma |U|^2U=0,
\end{equation}
where $\sigma=\pm 1$ denotes self-focusing and self-defocusing
nonlinearity, and $n(x)$ is a 1D lattice. For definiteness, we take
\[ n(x)=-6\sin^2x \]
in this paper (see Fig. \ref{f:fig1}(a)). Stripe (1D) solitons in
this model are of the form $U(x,y,z)=u(x)e^{-i\mu z}$, where $u(x)$
is a real-valued localized function, and $\mu$ is the propagation
constant. When $\sigma=-1$ (defocusing nonlinearity), there is a
stripe-soliton family in the first bandgap of the lattice. The power
curve of this family is shown in Fig. \ref{f:fig1}(b). Here, the
power $P$ is defined as $\int_{-\infty}^\infty |u|^2 dx$. The
intensity profile of the soliton at $\mu=4.5$ is displayed in Fig.
\ref{f:fig1}(c) (the peak intensity is approximately 3.2). To
determine the transverse stability of these stripe solitons, we
perturb them as
\begin{eqnarray*}
U(x,y,z)&=& e^{-i\mu z}\left\{u(x)+[v(x)+w(x)]e^{iky+\lambda z}
\right.   \hspace{1cm} \\
&& \hspace{1cm} \left. +[v^*(x)-w^*(x)]e^{-iky+\lambda^*z}\right\},
\end{eqnarray*}
where $v, w\ll 1$ are normal-mode perturbations, and $k$ is the
transverse wavenumber. Substituting this perturbed solution into
(\ref{e:model}) and neglecting higher order terms of $(v, w)$, we
obtain the linear-stability eigenvalue problem
\begin{equation*}
L_0w=-i\lambda v, \qquad L_1v=-i\lambda w,
\end{equation*}
where
\begin{eqnarray}
L_0& = &\partial_{xx} +n(x)+\mu -k^2 +\sigma u^2,  \nonumber \\
L_1& = &\partial_{xx} +n(x)+\mu -k^2 +3\sigma u^2, \nonumber
\end{eqnarray}
and $\lambda$ is the eigenvalue. The full spectrum of this
eigenvalue problem can be obtained numerically by the Fourier
collocation method \cite{Yang_SIAM}. For the stripe soliton at
$\mu=4.5$ (see Fig. \ref{f:fig1}(c)), its stability spectrum is
displayed in Fig. \ref{f:fig1}(e). This spectrum lies entirely on
the imaginary axis, indicating that this stripe soliton is
transversely stable under defocusing nonlinearity! In contrast, when
the nonlinearity is self-focusing ($\sigma=1$), stripe solitons will
remain transversely unstable, and this instability is strong. To
demonstrate, a family of stripe solitons in the semi-infinite gap
under focusing nonlinearity is obtained, and its power curve is
plotted in Fig. \ref{f:fig1}(b). At $\mu=1$ of the power curve, the
soliton is shown in Fig. \ref{f:fig1}(d) (its peak intensity is
roughly 1.7). The linear-stability spectrum of this soliton is
displayed in Fig. \ref{f:fig1}(f). This spectrum contains large
positive eigenvalues (with maximum 1.25), indicating strong
instability.

Next we consider the linear stability of other stripe solitons in
the solution families of Fig. \ref{f:fig1}(b). When the solitons are
near an edge $\mu_0$ of a Bloch band, these solitons under
perturbations are low-amplitude Bloch-wave packets,
\begin{equation*}
U(x,y,z)=e^{-i\mu_0z}\left[\epsilon \Psi(X,Y,Z) p(x)+\epsilon^2 U_2
+ \cdots\right],
\end{equation*}
where $p(x)$ is the Bloch wave at edge $\mu_0$, $0<\epsilon\ll 1$,
$X=\epsilon x$, $Y=\epsilon y$, $Z=\epsilon^2z$, and $\Psi(X,Y,Z)$
satisfies
\begin{equation}  \label{e:Psi}
i\Psi_Z+D\Psi_{XX}+\Psi_{YY}+\sigma\alpha |\Psi|^2\Psi=0,
\end{equation}
with $D$ being the diffraction coefficient of the 1D lattice at edge
$\mu_0$ and $\alpha>0$ being a constant \cite{Yang_SIAM}. Stripe
solitons in Eq. (\ref{e:model}) correspond to stripe envelope
solutions $\Psi(X,Y,Z)=A(X)e^{-i\tau Z}$, where $\mbox{sgn}(\tau)
=-\mbox{sgn}(\sigma)=-\mbox{sgn}(D)$, and $A(X)$ is a sech function
\cite{Yang_SIAM}. It is well known that this stripe envelope
solution is transversely unstable in the envelope equation
(\ref{e:Psi}) \cite{Zak_Rub,Kiv_Peli,Yang_SIAM}. Thus low-amplitude
soliton stripes in Eq. (\ref{e:model}) are all transversely
unstable. In particular, for the stripe-soliton family in the
semi-infinite gap, $D>0$, thus the transverse instability is of
neck-type (due to positive eigenvalues); while for the soliton
family in the first gap, $D<0$, thus the transverse instability is
of snake-type (due to both positive and complex eigenvalues)
\cite{Yang_SIAM}. The magnitude of these unstable eigenvalues is
proportional to $\sqrt{|\mu-\mu_0|}$ \cite{2Dtrain_stable}.

Away from band edges, we have tracked these transverse-instability
eigenvalues for the two soliton families in Fig. \ref{f:fig1}(b). We
find that for the soliton family in the semi-infinite gap (under
focusing nonlinearity), as $\mu$ moves away (decreases) from the
band edge $\mu_0=2.06$, the magnitude of the largest positive
eigenvalue keeps increasing (at $\mu=1$, this magnitude has reached
1.25, see Fig. \ref{f:fig1}(f)). Thus all stripe solitons in this
family are transversely unstable. This instability is of neck-type
and is strong when $\mu$ is not near the band edge.

\begin{figure}[h]
\center
\includegraphics[width=0.42\textwidth]{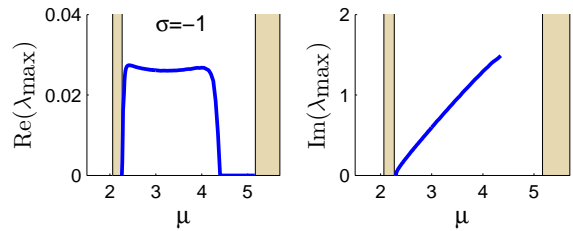}

\vspace{-0.2cm} \caption{The most unstable eigenvalue
$\lambda_{\mbox{max}}$ versus $\mu$ for stripe solitons in the first
gap under defocusing nonlinearity: (left) real part; (right)
imaginary part. }  \label{f:fig2}
\end{figure}

However, for the soliton family in the first gap under defocusing
nonlinearity, the story is very different. In this case, as $\mu$
moves away (increases) from the band edge $\mu_0=2.27$, the positive
eigenvalues of snake instability quickly disappear when $\mu > 2.3$.
Meanwhile, the real parts of the complex eigenvalues in the snake
instability first saturate quickly to a very low level when
$2.4<\mu<4.3$. After $\mu>4.3$, these complex eigenvalues then
totally disappear, hence stripe solitons in this $\mu$ range are all
stable! To demonstrate, the most unstable eigenvalue
$\lambda_{\mbox{max}}$ versus $\mu$ is plotted in Fig. \ref{f:fig2}.
It is seen that the real part of $\lambda_{\mbox{max}}$ (the maximum
growth rate) is under 0.03 for all stripe solitons, indicating that
the transverse instability is extremely weak. In addition, when
$\mu>4.3$, the real part of $\lambda_{\mbox{max}}$ becomes zero,
hence those stripe solitons are linearly stable. These results show
that under defocusing nonlinearity, transverse instability of stripe
solitons is either greatly suppressed or totally eliminated.

\begin{figure}[h]
\includegraphics[width=0.48\textwidth]{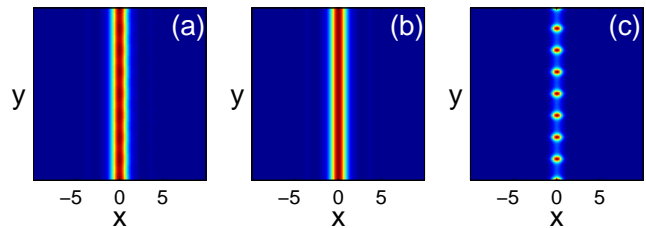}

\vspace{-0.2cm} \caption{(a) Initial intensity pattern of the stripe
soliton in Fig. \ref{f:fig1}(c) under 2\% perturbations; (b) output
intensity of the perturbed soliton in (a) after nonlinear evolution
of $z=100$ under defocusing nonlinearity; (c) output intensity of
the perturbed soliton in Fig. \ref{f:fig1}(d) after nonlinear
evolution of $z=3$ under focusing nonlinearity. } \label{f:fig3}
\end{figure}

The above linear-stability results are also corroborated by
nonlinear-evolution simulations of these stripe solitons under
random-noise perturbations. For the soliton in Fig. \ref{f:fig1}(c),
its initial perturbed state (with 2\% random-noise perturbations) is
shown in Fig. \ref{f:fig3}(a), and evolution output of this
perturbed soliton under defocusing nonlinearity at $z=100$ is shown
in Fig. \ref{f:fig3}(b). It is seen that even after such a long
evolution, this stripe soliton still remains robust and does not
break up. In contrast, when the soliton in Fig. \ref{f:fig1}(d) is
perturbed by the same amount of perturbations, after evolution under
focusing nonlinearity, this stripe quickly breaks up into filaments
at $z=3$ (see Fig. \ref{f:fig3}(c)). Thus suppression of transverse
instability under defocusing nonlinearity and persistence of
transverse instability under focusing nonlinearity hold for
nonlinear evolutions as well.

\begin{figure}
\includegraphics[width=0.48\textwidth]{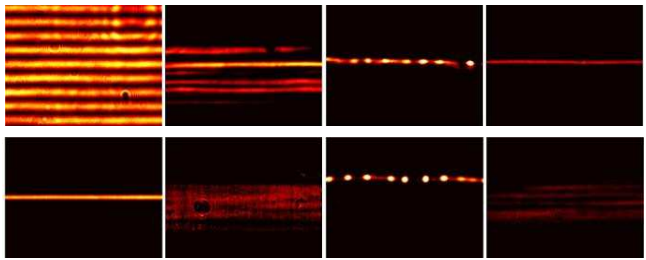}

\vspace{-0.1cm} \caption{Experimental results. Upper first column:
1D lattice; lower first column: initial probe beam; upper row:
results with lattice; lower row: results without lattice; second
column: linear diffraction; third column: output with focusing
nonlinearity; fourth column: output with defocusing nonlinearity.
The bias field and crystalline c-axis are along the vertical
direction.} \label{f:experiment}
\end{figure}

Experimentally, we have confirmed the above theoretical predictions.
The experiments were performed in a 10mm-long biased SBN:60
photorefractive crystal. The optically induced 1D lattice (41$\mu$m
lattice spacing) is shown in the upper first panel of Fig.
\ref{f:experiment}, while the initial stripe beam (12$\mu$m FWHM) is
shown in the lower first panel. The peak intensity ratio between the
probe and lattice beams is about 1:5. In the presence of the
lattice, this probe beam exhibits strong discrete diffraction during
linear propagation (upper second panel). It breaks up due to strong
transverse instability under self-focusing nonlinearity at a bias
field of 2kV/cm (upper third panel), but remains robust against
transverse perturbations under defocusing nonlinearity at a bias
field of $-1.6$kV/cm (upper fourth panel). As a comparison, when no
lattice is induced inside the crystal, the corresponding results are
shown in the lower panels of Fig. \ref{f:experiment}. In this case,
the stripe beam breaks up strongly due to neck-type transverse
instability under focusing nonlinearity \cite{Saffman}, and broadens
even more than linear diffraction under defocusing nonlinearity.

\begin{figure}[h]
\includegraphics[width=0.48\textwidth]{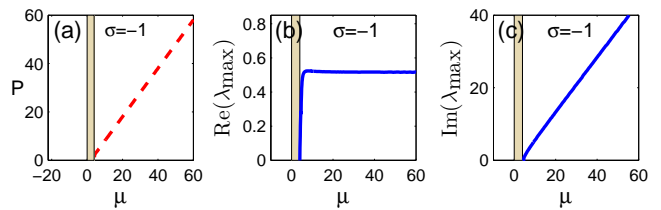}

\vspace{-0.3cm} \caption{(a) Power curve of discrete stripe solitons
under defocusing nonlinearity; (b,c) real and imaginary parts of the
most unstable eigenvalue $\lambda_{\mbox{max}}$ versus $\mu$.}
\label{f:fig5}
\end{figure}

It is important to notice from Fig. \ref{f:fig1}(b) and Fig.
\ref{f:fig2} that transversely stable stripe solitons (under
defocusing nonlinearity) are located near the second Bloch band,
implying that their stability is caused by mode coupling between the
first and second bands. This means the existence of these stable
stripe solitons cannot be predicted by the corresponding discrete
NLS model,
\begin{equation*}
iU_{n,z}+U_{n+1}-2U_n+U_{n-1}+U_{n,yy}+\sigma |U_n|^2U_n=0,
\end{equation*}
since this discrete model is derived under a single-band
approximation and it does not incorporate mode coupling between
different Bloch bands \cite{Panos_book}. To confirm this, we seek
stripe solitons in this discrete model as $U_n(y,z)=u_n e^{-i\mu
z}$, where $\mu$ is the propagation constant. The power curve of
these solitons $P=\sum |u_n|^2$ versus $\mu$ for $\sigma=-1$
(defocusing nonlinearity) is plotted in Fig. \ref{f:fig5}(a), and
the most unstable linear-stability eigenvalue of each soliton versus
$\mu$ is plotted in Fig. \ref{f:fig5}(b,c). It is seen that when
$\mu$ moves away (increases) from the band edge $\mu_0=4$, the real
part of the most unstable eigenvalue quickly saturates (to about
0.52) and then stays at this level for all higher $\mu$, thus
transverse instability persists for all these discrete stripe
solitons under defocusing nonlinearity.

%Physically, this suppression of transverse instability of stripe
%beams in a 1D lattice under defocusing nonlinearity can be
%heuristically understood. In this case, the repeated Bragg
%reflection of the lattice causes the stripe beam to be horizontally
%localized. Transversely, the instability is absent since the
%nonlinearity is self-defocusing.

In summary, we have demonstrated both theoretically and
experimentally that the transverse instability of coherent soliton
stripes is totally eliminated or greatly suppressed when the soliton
stripes propagate in a 1D photonic lattice under self-defocusing
nonlinearity. This elimination of transverse instability makes
stripe solitons applicable in physical settings.

This work was supported by the NSF and AFOSR.

\end{document}